# Creative Foraging: A Quantitative Paradigm for Studying Creative Exploration

Yuval Hart[1,2], Avraham E Mayo[1,2], Ruth Mayo[3], Liron Rozenkrantz[4], Avichai Tendler[1,2], Uri Alon[1,2]* and Lior Noy[1,2]*

1 Department of Molecular Cell Biology, Weizmann Institute of Science, Rehovot, Israel

2 The Theatre Lab, Weizmann Institute of Science, Rehovot, Israel

3 Department of Psychology, The Hebrew University of Jerusalem, Jerusalem, Israel

4 Department of Neurobiology, Weizmann Institute of Science, Rehovot, Israel

* Corresponding author

## Abstract

Creative exploration is central to science, art and cognitive development. However, research on creative exploration is limited by a lack of high-resolution automated paradigms. To address this, we present such an automated paradigm, the creative foraging game, in which people search for novel and valuable solutions in a large and well-defined space made of all possible shapes made of ten connected squares. Players discovered shape categories such as digits, letters, and airplanes. They exploited each category, then dropped it to explore once again, and so on. Aligned with a prediction of optimal foraging theory (OFT) prediction, during exploration phases, people moved along meandering paths that are about three times longer than the minimal paths between shapes; when exploiting a category of related shapes, they moved along the minimal paths. The moment of discovery of a new category was usually done at a non-prototypical and ambiguous shape, which can serve as an experimental proxy for creative leaps. People showed individual differences in their search patterns, along a continuum between two strategies: a mercurial quick-to-discover/quick-to-drop strategy and a thorough slow-to-discover/slow-to-drop strategy. Contrary to optimal foraging theory, players leave exploitation to explore again far before categories are depleted. This paradigm opens the way for automated high-resolution study of creative exploration.

## Introduction

Creative exploration and discovery processes are defined as the search for novel and valuable elements within a set of constraints. Creative exploration is central for artists (Kozbelt, 2006), designers (Dorst & Cross, 2001) and scientists (Klahr & Simon, 1999). More generally, it is an integral part of cognitive development (Buchsbaum, Bridgers, Skolnick Weisberg, & Gopnik, 2012; Gelman & Gottfried, 2006). Despite progress achieved in understanding creative search processes from different perspectives (Amabile & M., 1996; Bowden, Jung-Beeman, Fleck, & Kounios, 2005; Dietrich & Kanso, 2010; Guilford, 1967; J. C. Kaufman & Sternberg, 2010; Mednick, 1962; Runco, 2007; Sternberg & Davidson, 1995; Weinberger, Iyer, & Green, 2016), much is still unknown about their underlying dynamics and mechanisms.

One obstacle for understanding creative exploration is the lack of automated and high-resolution experimental paradigms. For example, one of the most commonly used tests for creativity is the alternate usage task (AUT)(Wilson, Guilford, & Christensen, 1953). In AUT participants are asked to come up with creative uses for common objects (e.g., a ping-pong ball). Participants show exploration patterns such as switching between categories of solutions (Runco, 2007). This test is coded manually, a laborious process which limits the extent of feasible studies. The test is difficult to analyze and model because the space of possible solutions is undefined (how many different usages for a ping-pong ball exist?) and has no natural notion of similarity between different solutions. Other common creativity tests share these limitations (Torrance, 1968).

More fundamentally, AUT and other common creativity tests record only the discovered solutions and not the intermediate steps leading from one solution to another. Thus, they do not allow insight into the process of exploration. Several studies attempted to map the processes leading to creative solutions by recording verbal transcripts of thought processes - the 'think aloud' method (Dorst & Cross, 2001; Kaplan & Simon, 1990). However, these methods are difficult to code and quantitatively analyze, and hence have a low throughput of measurements.

Here, we introduce a new paradigm for creative exploration that allows automated, high resolution tracking of the search process in a space of possibilities that is large yet well defined. To achieve this, we borrow techniques from the study of human foraging.

In a series of studies Hills, Todd and colleagues used a spatial foraging task and a letter puzzle task to experimentally study human foraging behavior (Hills, Todd, & Goldstone, 2008; Wilke, Hutchinson, Todd, & Czienskowski, 2009; Hills, Jones, & Todd, 2012; Hills, Todd, & Goldstone, 2015). These experimental foraging tasks employed solution spaces which are finite and well-defined(Newell & Simon, 1972). For example, in the spatial foraging task participants searched for resources on a grid of locations. This task records intermediate steps between solutions and has a clear distance metric, features that allow the foraging behavior to be quantitatively analyzed and mathematically modeled using optimal foraging theory (OFT)(Charnov, 1976).

In order to develop a similar task for creative search, we need to address one key difference between foraging and creative search. In the foraging tasks, the goal is to find as many solutions as possible; in creativity tests the goal is two-fold: to find many solutions (fluency) of high quality (novel and valuable). In other words, whereas the value of the different solutions in the foraging tasks is constant (for example, all foraged pixels in the spatial foraging task are equally valuable), the distribution of solution values in creativity tests is skewed, with many low-quality solutions (boring or non-feasible) and only a handful of high-quality creative solutions.

Here we present a paradigm for creative foraging in which players seek solutions that they find interesting and beautiful, embedded within a well-defined search space. The *creative foraging game* enables high resolution measurement of creative, intrinsically motivated search in the framework of visual exploration and discovery(Finke, 1990; Ward, 1994; Johnson-Laird, 2004; Jennings, Simonton, & Palmer, 2011; Noy et al., 2012).

The creative foraging game is a computer game in which players can move ten adjacent squares to create different shapes (Fig. 1, SI, Fig. S1). There are 36,446 possible shapes (SI, Section S1). The players are instructed to save shapes that they find interesting and beautiful into a gallery. The timing of all moves and gallery choices is recorded. This setting allows us to observe all intermediate steps as people search a large but finite metric space of different possibilities.

We present data on 100 participants and interpret the results in the context of OFT according to the following hypotheses:

($H_1$) Participants' trajectories in shape space are composed of distinguishable phases of exploration and exploitation: exploration for new shape categories and exploitation of similar shapes within a category.

($H_2$) In phases of exploitation people move along optimal (shortest) paths.

Our third hypothesis diverges from the predictions of OFT. The performance metric in OFT is the amount of collected resources. Hence OFT predicts that exploitation of a given patch is terminated due to diminishing returns caused by depletion of the patch. In contrast, in creative foraging the intrinsic interest 'value' of collected resources is important. Collection from a given patch may lead to diminishing interest, far before the patch is depleted. We therefore hypothesize that:

($H_3$) In creative foraging participants leave exploitation phases sooner than predicted by OFT, far before patch depletion.

The next hypothesis concerns the moment when a new category of shapes is discovered. We suggest that entry points to phases of exploitation can serve as an experimental proxy for creative leaps. Qualitative descriptions of creative processes suggest that creative leaps often occur in the periphery of a field (Gardner, 2011), through a road that is not often taken. In addition, creativity was previously linked to ambiguity (Rothenberg, 1986). We therefore hypothesize that:

($H_4$) Shapes at the entry points to phases of exploitation are more ambiguous, in the sense that they can lead different players to different categories.

Finally, we hypothesize that

($H_5$) People show individual differences in their search strategies and patterns.

In the results section we operationalize these hypotheses in terms of the creative foraging task.

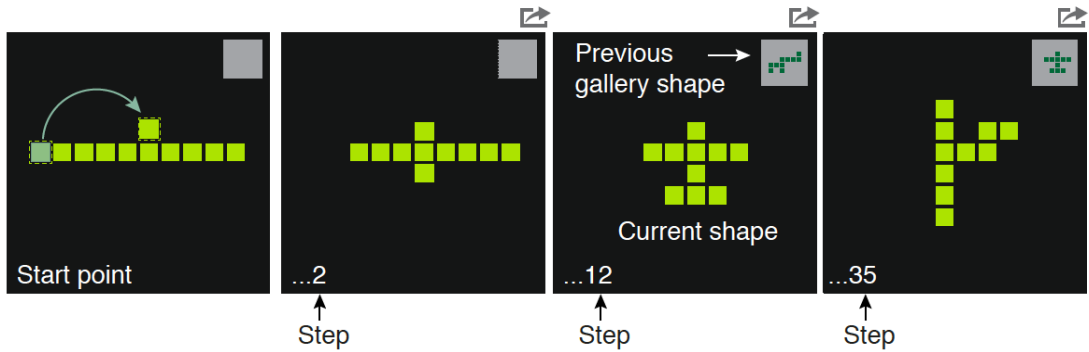

**Figure 1: The creative foraging game allows high-resolution automated analysis of exploration and discovery processes.** Players move squares starting from a 10-square line with the aim of finding 'interesting and beautiful' shapes. Moves keep all squares connected (not by a diagonal). A shape can be saved to a gallery by pressing the gray square at the top-right corner of the screen; in this square, the last shape chosen to the gallery is displayed.

## Results

### Search dynamics are composed of alternating phases of exploration and exploitation

We analyzed games by 100 participants. Games lasted 15 min and averaged 306 steps (95% CI=[286,324]) and 46 shapes chosen to the gallery (95% CI=[41,50]). We analyzed the time series of gallery choices made by each player. The timing difference between subsequent gallery choices shows a distinct pattern: periods of slowing down where gallery choices happens more and more infrequently, and periods of acceleration where gallery choices occur more and more rapidly. We used a thresholding algorithm (Methods, SI, Section S3) that employs this timing data to segment the games into two alternating phases, which we name exploration and exploitation respectively (See Fig. 2). Games showed a median of 7 exploration and exploitation phases (95% CI=[6,8]). Properties of the two phases are shown in table 1.

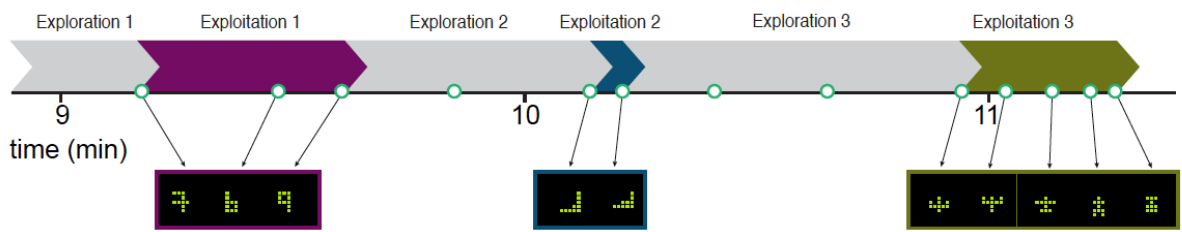

**Figure 2: Search trajectories are segmented into exploration and exploitation phases.** Segmentation uses the timing difference Δt between gallery shape choices. Periods with increasing Δt are labelled as exploration phases. An exploration phase ends with a decrease in Δt, leading to a sequence of rapid gallery choices, labelled as exploitation phases (see Methods, SI, Section S3). Choices of shapes to the gallery are marked with an open circle.

**Table 1 Characteristics of exploration and exploitation phases (Medians and 95% CI)**

|  | Exploration | | Exploitation | | |
|---|---|---|---|---|---|
|  | Median | 95% CI | Median | 95% CI |  |
| Number of gallery shapes per phase | 1.3 | [1.28,1.34] | 3.2 | [3,3.45] | Exploration < Exploitation |
| Number of moves between gallery shape | 10 | [9,11] | 4.4 | [4,4.8] | Exploration > Exploitation |
| Duration of phase (sec) | 58 | [52,64] | 52 | [47,58] | Same |
| Number of moves per phase | 12.5 | [10,14] | 12.3 | [11,15] | Same |
| Duration of a single move (sec) | 3 | [2.8,3.4] | 2.9 | [2.7,3] | Same |

**In exploitation phase players quickly harvest a sequence of similar shapes**
We noticed that the gallery shapes gathered in a given exploitation phase were similar to each other. To test this, we performed an 'odd-shape-out' test with a new set of participants who did not play the game (Fig. 3). We find that people can differentiate between three gallery shapes taken from the same exploitation phase and a distractor gallery shape taken from a different exploitation phase ($\chi 2(2, N=67) = 1627$, $p < .0001$, SI, Section S4). We conclude that players harvest a sequence of similar of shapes in each exploitation phase. They then leave this 'patch' of shapes to explore for new shapes ($H_1$).

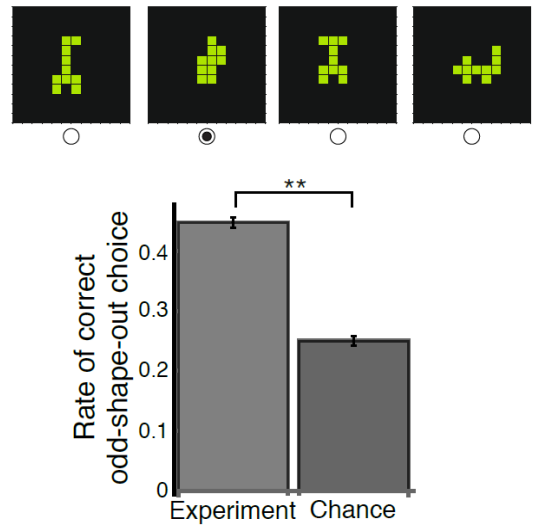

**Figure 3: Shapes found in an exploitation phase are perceived as similar.** Participants who did not play the creative foraging game chose the odd-one-out among four shapes, three from the same exploitation phase and one from a different exploitation phase found by the same player. Participants chose the outlier twice as often as by chance. Means and STD (computed using bootstrapping) are shown ($p<0.0001$, see SI, Section S4).

**Only in exploitation phases players move along direct paths between gallery shapes**

Next, we considered a prediction of optimal foraging theory (OFT) that within exploitation phases, players move along direct (optimal, shortest) paths between discoveries (Pyke, 1984). To test this, we compared the length of players' actual paths between two gallery shapes to the length of the shortest possible path between these shapes. In doing so, we took advantage of the fact that the current search space is well-defined and fully enumerated, so that the shortest path between any two shapes can be readily computed.

We find that during exploration phases players did not take the shortest path between gallery shapes; their paths averaged three times longer than the shortest possible path (Median shortest distance/actual distance = 0.35, 95% CI = [0.3,0.4]) (Fig. 4). In contrast, in exploitation phases, players moved closer to the optimal path (Median shortest distance/actual distance = 0.74, 95% CI = [0.67,0.8], MW-test=985, $p < .001$, effect = 0.8). These effects are evident also when comparing pairs of shapes with the same shortest distance in the two phases (SI, Fig. S4). In summary, we find that in phases of exploitation – but not in phases of exploration - people move along optimal (shortest) paths ($H_2$).

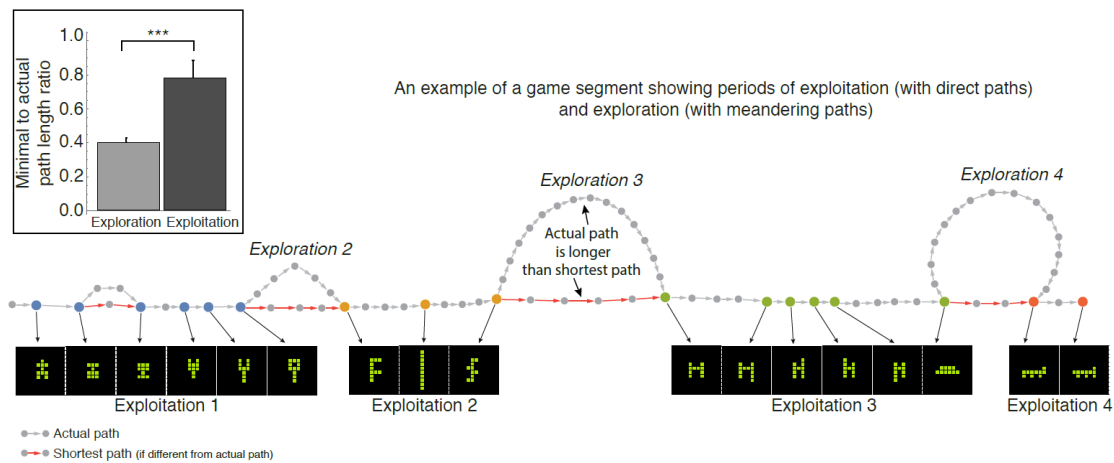

**Figure 4: Participants follow optimal (shortest) paths in exploitation phases.**
Shown are moves (circles), gallery-chosen shapes (colored circles), actual (gray) and minimal (red) paths (minimal path marked when different from actual path). Exploration and exploitation phases are labelled. Inset, players' median ratio of minimal to actual path length between gallery shapes in exploitation and exploration (Exploration: Median = 0.35, 95% CI = [0.3,0.4], Exploitation: Median = 0.74, 95% CI = [0.67,0.8], MW-test=985, p < .001, effect = 0.8).

**Different players find similar shape categories**

We noticed that different players rediscovered the same categories of shapes- for example numerical digits and shapes resembling airplanes. To study this, we developed an automated definition of shape categories. We defined a shape category as a set of shapes often rediscovered by different players. To do so, we used a network representation of shapes found by different players and used a community-finding algorithm to automatically detect shape categories (see Methods and SI, Section S6). The algorithm revealed 14 categories that were found again and again by different players. These 14 shape categories contain a total of 653 shapes, 34% of all gallery shapes collected by participants.

Each category includes shapes with a shared theme: visual similarity, familiar objects such as animals or airplanes, or symbols such as digits or letters. Examples of shape categories are shown in Fig. 5.

We tested whether people can reliably distinguish between these categories using an 'odd-shape-out' test, similar to the one reported above. We asked 86 people who did not play the game to match a random triplet of shapes from a single category either to a set of six shapes from the same category or to a set of six shapes from another random category (see SI, Section S7, Fig. S5B). Participants chose the correct set 80±1% of the times, significantly more than chance ($\chi2(2, N=86) = 1343$, p < .001).

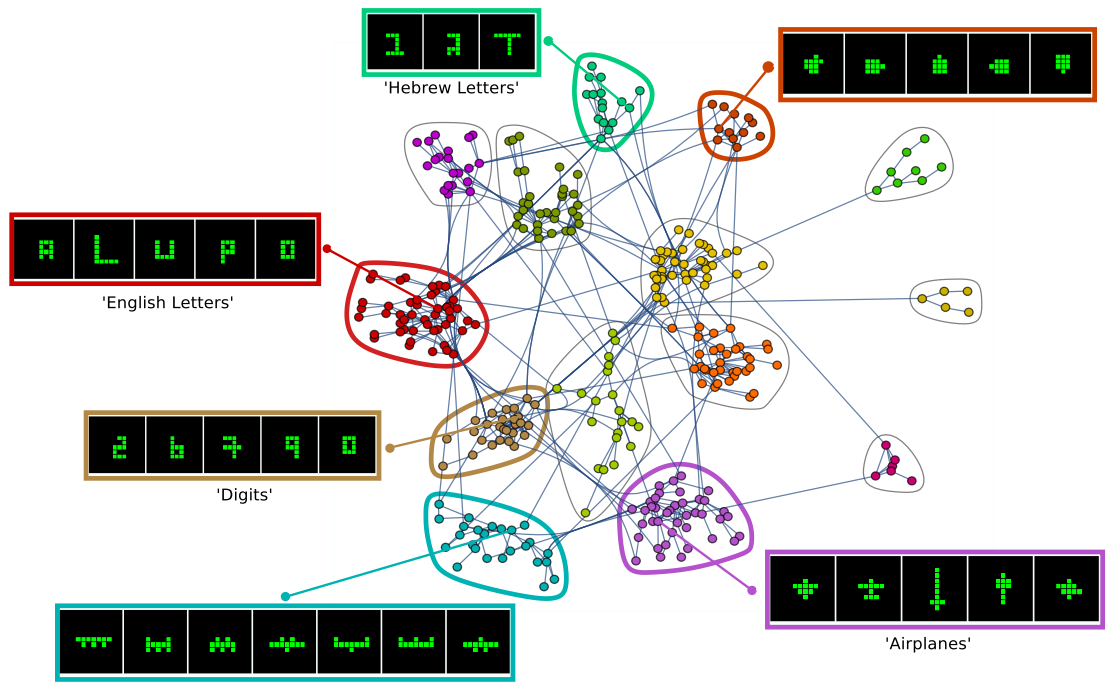

**Figure 5: Different players discover similar categories of shapes in their exploitation phases.** Network in which nodes are patches (shapes found in an exploitation phase) from 100 players. Links connect patches that share at least two shapes. Categories are defined as Girvan–Newman modules in the graph, and are shown in colors, with representative shapes (see SI, Section S6, Figs. S5-S6).

We conclude that the creative foraging game induces meaning on the space of shapes, with at least 14 different meaning categories. People explore until they hit upon one of these categories, then exploit that category, and then return to explore for new categories.

Interestingly, the shape categories are not separated from each other by many moves. Instead they are interleaved such that neighboring shapes belong to different categories (see Methods). Players in an exploitation phase seem to focus their attention on one category, and pass by many shapes from other categories.

**Departure from exploitation phase is not due to depleted resources**
Identifying the shape categories enabled us to test another prediction of optimal foraging. According to OFT, foragers leave a patch of resources when they begin to deplete it, a phenomenon known as diminishing returns. Diminishing returns are due to a decrease in the rate of harvesting resources. To test this we asked whether participants leave an exploitation phase when they are close or far from depleting the current category.
We find that players cover on average only 6.8% (95% CI=[6.1,7.4]) of each category (see SI, Section S9 and Fig. S8). Moreover, at the point of departure from an exploitation phase, the next potential shape belonging to the same category is very close, only 1.3 moves away on average (95% CI=[1.2,1.4]). Thus, players leave an exploitation phase far before the category is depleted ($H_3$).

**Players showed individual differences in their search along a continuum between fast and slow strategies**

We next asked about the difference in the search process of different people ($H_5$). Players varied considerably in the average duration of their exploitation phases (Median = 50 sec; 5%-95% range=[17,102] sec). We find that the duration of exploration and exploitation phases was positively correlated between participants: players with long exploration phases also had long exploitation phases (Spearman correlation, r=0.80, 95% CI=[0.69,0.87], p<0.001).

To adjust for possible differences in individual move speeds, we plotted the mean number of moves in exploration and exploitation for each player (Fig. 6). The average number of moves in exploration and exploitation phases also showed strong correlation (Spearman correlation, r=0.78, 95% CI=[0.68,0.86], p<0.001). We controlled for the possibilities that these effects are the result of our segmentation algorithm or the general distributions of the duration of each phase (SI, Section S10 and Fig. S9).

We conclude that people in our sample vary along a continuum between two strategies: a fast strategy of short exploration/exploitation and a slow strategy of long exploration/exploitation. Those quick to discover are quick to drop, and those slow to discover are thorough in their exploitation of a category.

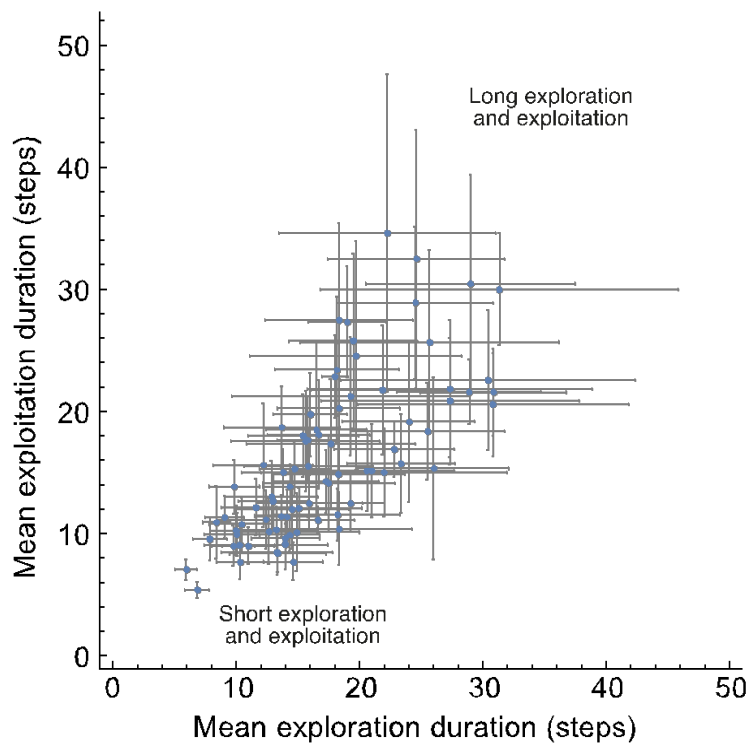

**Figure 6: Players show correlated mean exploration and exploitation phase**

**durations.** Mean and error bars of one std are shown for each player. Spearman correlation, r=0.78, 95% CI=[0.68,0.86], p<0.001.

### Entry into exploitation phases occur at ambiguous transition shapes at the periphery of shape categories

The creative foraging game allowed us to focus on the moments of discovery of new categories of shapes. These moments occur in the transition between exploration and exploitation. We call the shape that marks the beginning of a new exploitation phase the 'transition shape'. Different players enter exploitation of a given category through different transition shapes. For example, in Fig. 7A, different players discover the 'digits' category through different transition shapes. One player entered the category of digits through a shape which we termed 'the 9 which is not a 9' (Fig. 7A, first row), whereas another player entered the digits categories through 'the 4 which is not a 4', Fig. 7A, second row). After finding the transition shape, players go on to find more prototypical digit shapes (Fig. 7A).

Only 23% (95% CI=[20, 25]) of the players who found a transition shape use it as a start of a new exploitation phase; for other players who reached the same shape, it represents a 'road not taken' (SI, Sections S11 and Fig. S10). Of those that did go on to start an exploitation phase, the category exploited differs between players. In this sense, transition shapes are ambiguous ($H_4$): they belong to multiple meaning categories more often than other gallery shapes (Fig. 7B,C) (transition shapes: Median = 50%, 95% CI=[49, 50]; non-transition shapes: Median = 15%, 95% CI=[15,16], SI, Section S12).

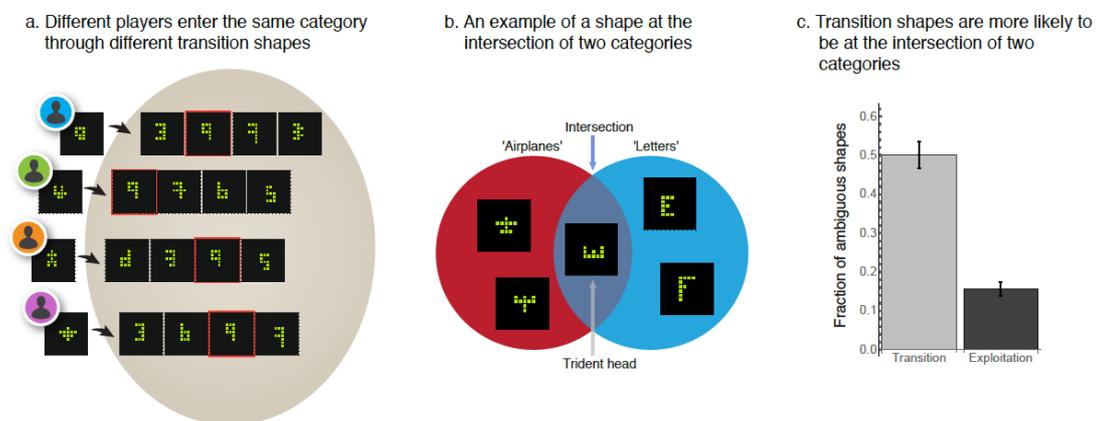

**Figure 7: Discovery of a new exploitation phase occurs at a transition shape that is ambiguous in the sense that it belongs to multiple categories. A)** Examples of transition shapes at the entry to the category of 'digits' by four different players. Transition shapes are non-prototypical, for example a 'four that is not a four', whereas shapes found after the transition shape tend to be more prototypical digits. **B)** A shape is ambiguous if it lies in the intersection of two or more categories (signifying at least two contexts of meaning), as exemplified by the trident shape shared by the categories of 'airplanes' and 'English letters'. **C)** Transition shapes are more likely to be ambiguous shapes than non-transition gallery shapes (transition shapes: Median = 50%, 95%

CI=[49, 50]; non-transition shapes: Median = 15, 95% CI=[15,16]).

**The creative foraging game can score creativity and correlates well with the manual AUT**

Finally, we compared the present test to a commonly used creativity test, the alternate uses test (AUT) (Wilson et al., 1953). The AUT provides two creativity measures: fluency- the total number of alternative usages found, and originality- the rarity of the solutions compared to a given dataset. We compared these measures to the corresponding measures from the creative foraging game, in a separate experiment on 90 people who played the CFG and took the AUT (SI, Section S13). Fluency in the CFG was defined as the total number of gallery shapes found by the player; originality was defined by the frequency that the exploitation gallery shapes of the player were also found by the 100 players in our main dataset.

We find that fluency and originality were correlated in both the AUT (Spearman correlation r=0.81, p < 0.001) and CFG (r = 0.29, p = 0.03), and thus a composite score can average over the two measures. To define a composite creativity score we averaged the Z-transformed fluency and uniqueness score of each player. We find that the composite creativity score in the CGF shows medium correlation with the composite creativity score in AUT (N=57, r = 0.3, p = 0.02, CI = [0.1, 0.5]). We conclude that the CFG is a satisfactory automated test for scoring the type of creativity tested by the AUT.

**Discussion**

We presented the creative foraging game as a paradigm to study intrinsically motivated creative exploration. In the CGF people search a defined metric space of shapes for interesting and beautiful shapes. Multi-dimensional information about the players search trajectories is automatically recorded at high resolution.

We find that participants' trajectories are composed of alternating phases of meandering exploration, punctuated by transitions to exploitation of patches of similar shapes ($H_1$). Different people rediscover the same shape categories, including letters, digits, airplanes and other categories with shared visual properties. Thus, the creative foraging game induces meaning categories on the space of shapes. We find that these categories are interleaved rather than segregated in this space, such that shapes from a given category have neighbors from different categories. Hence, participants need only change their focus of attention in order to discover a new category. This contrasts with a common metaphor in which creative search is like seeking an oasis of novel and valuable solutions in the desert of ordinary. Search in an interleaved landscape might be relevant for exploring creative solutions in information-rich digital environments.

The existence of exploration and exploitation phases suggests a link to studies on human foraging behavior. Human foraging studies supported the predictions of OFT (Hills et al., 2008, 2012). In line with OFT, we find that in exploitation - but not in exploration - participants follow the optimal path (shortest possible path) between the collected shapes ($H_2$). This finding of meandering paths in exploration versus direct paths in exploitation can ground the metaphors from creativity research of 'fogginess' in exploration vs. 'visibility' in exploitation(Jennings et al., 2011; Simonton, 2003).

Moreover, the distinct behavioral patterns of exploitation and exploration measured in the current work raise the possibility that they could correspond to changes in brain activation that have been associated with exploration and exploitation states (Aston-Jones & Cohen, 2005).

Our results differ from OFT regarding the point of departure from an exploitation phase. OFT assumes that foragers leave a patch due to diminishing returns caused by resource depletion. In contrast, we find that participants exit exploitation far before their patch of shapes is depleted ($H_3$). This difference between regular and creative foraging might be the result of different goals. Regular foraging aims to maximize the amount of discovered solutions, where all solutions have the same quality. In contrast, in creative foraging the intrinsically evaluated 'quality' of solutions varies considerably - some solutions are novel and interesting while others are more dull. In the current paradigm, the process of evaluating solution quality might be dynamic: the value of the first exemplar from a category is higher than the value of the $n^{th}$ exemplar, suggesting that creative foraging is guided by a mechanism related to novelty seeking (Dahan, Noy, Hart, Mayo, & Alon, 2016; Dellu, Piazza, Mayo, Le Moal, & Simon, 2008).

We also studied individual differences in search strategies. We find strong correlation in the length of exploration and exploitation phases: while some people tend to explore for a short time and exploit for a short time (a mercurial strategy of quick to find, quick to drop), other explore longer and exploit longer (a thorough strategy of slow to find, slow to drop). This correlation is preserved also when we adjusted for the speed in which people make the moves in the game. Other possible strategies such as short exploration and long exploitation (quick to discover and thorough exploitation) seem to be missing in our sample. More subtle differences in exploration and exploitation strategies between individuals might be observed with a larger pool of participants. Future research can explore possible mechanisms in terms of different personality traits[35] or attention deficits[36], and to link these findings with the changes in brain activation that have been associated with exploration and exploitation states (Aston-Jones & Cohen, 2005). It is encouraging that the present test correlates reasonably well with the arguably the most commonly used manual test for creativity, the AUT.

Finally, we characterized the moments of discovery of new categories using the shapes at which participants enter a new exploitation phase. We find that the transition between exploration to exploitation is done through shapes that are more ambiguous (belong to more than one category, $H_4$) than other gallery shapes. We suggest that these transition shapes can serve as an experimental proxy for moments of insight (Koestler, 1964).

The present study has several limitations: first, the participant pool is relatively small and from a single culture. This limitation can be readily overcome in future studies due to the suitability of the creative foraging game for online platforms. Second, participant creativity is impacted by the limited space of geometric shapes we ask them to explore. However, the effect of limitations on creative exploration is not clear, and some research suggests that certain constraints can enhance creativity(Finke, 1990; Haught-Tromp, 2016). Finally, our paradigm addresses only a thin slice of creative exploration (Green,

2016), and can be categorized as a study on small-c creativity (C. J. Kaufman & Beghetto, 2009).

Keeping these limitations in mind, the present paradigm can boost creative exploration research by providing automated information on how people explore, including intermediate steps, in a situation where solutions can be compared and their distance computed. This high-resolution approach revealed individual search patterns along a continuum between two strategies: quick-to-discover-quick-to-drop, and slow-to-discover-yet thorough, with other strategies such as combined quickness and thoroughness not found in our sample. The creative foraging game may thus open a window to study mechanisms, traits and interventions that can improve creative search.

## Methods

### Participants

Undergraduate students at the Hebrew university (54 females, 46 males, age 20-49, mean(±std) =25(±4)), took part in the experiment either for credit or payment equivalent to 5$.

### Creative foraging game

The creative foraging game was run on a PC with on-screen instructions. Players created shapes by moving at each step one of ten identical squares, keeping the squares connected by an edge (and not a corner, see SI, Fig . S1). The initial condition was ten squares in a horizontal line.  Participants were given the following instructions: *"Your goal is to explore the world of shifting shapes and discover those that you consider interesting and beautiful"*. At each point in the game players could store the current shape to a gallery by pressing a gray square at the top-right side of the screen. The gray square showed the last gallery shape chosen. The gallery had no limit on the number of shapes. Games lasted 15 minutes. After the game, players performed another task - choosing the five most creative shapes from their gallery. This task is not analyzed in the current study.

### Links to the creative foraging game and data

The game can be accessed at:
http://www.weizmann.ac.il/mcb/UriAlon1/Cubes/welcome.html

The raw data used in the analysis described in the main text can be accessed at:
http://www.weizmann.ac.il/mcb/UriAlon/download/downloadable-data

### Segmentation algorithm

A segmentation algorithm defined the exploitation and exploration phases in each game. The input is the series of timing difference in seconds between consecutive choices of gallery shapes. The output is a labeling of gallery shapes into exploration and exploitation phases (SI, Fig. S2). The algorithm has two iterations. In the first iteration, consecutive shapes are grouped together into an exploitation phase if their timing differences monotonically decrease. The second iteration groups together consecutive exploitation groups from the first iteration if the maximal timing difference in the earlier group is larger than the maximal time difference in the second group. The purpose of the second iteration is to avoid fragmentation of exploitation phases due to a single large time-difference value (see SI, Section S3 for further details). Shapes that remain outside of exploitation phases are labeled as exploration phases.

**Odd-shape-out test of perceived shape similarity**

We tested the similarity of shapes within the same exploitation phase by means of an odd-out test. Participants who did not play the creative foraging game (31 males, 36 females; age range: 20-80, mean (±std) = 36 (±12)) observed one of four blocks of 50 quartets of shapes. Sample size was calculated assuming medium effect size, and was sufficient for statistical power exceeding 0.95 assuming mu0=0.25, mu1=0.4, sigma=0.05, alpha=0.05. Each quartet included three shapes from the same exploitation phase and a fourth shape from a different exploitation phase found by the same player, in randomized order (see Fig. 3). A total of 200 randomly selected exploitation phases were tested out of 791 in the data-set (~25%). The experiment was performed on a commercial platform for online surveys (Qualtrics, see: http://qualtrics.com/). Error bars were computed by bootstrapping 10,000 times. See SI, Section S4 for further details.

**Shape Categories**

We defined categories of shapes found by different players in exploitation phases using a network community approach. We constructed an undirected network in which each node is a group of shapes found in a single exploitation phase, hereafter termed patch. Two patches were connected in the network if they share at least two shapes. The network had a giant component of 334 patches and 17 smaller connected components (containing less than 8 patches each with a total of 46). We defined shape categories by finding communities (modules) in the giant component using the Girvan–Newman algorithm. We find 14 shape categories of varying sizes (mean number of shapes (±std) = 70 (±38)). See SI, Table S1 for the characteristics of the shape categories and SI, Fig. S5 for all the unique shapes in the 4 biggest categories.

**Perceptual test of within-category similarity**

We tested the perceptual grouping of the categories by a grouping experiment. Participants who did not play the game (43 males, 43 females; age range: 18-77, mean (±std) = 41.6 (±15.2)) were asked to match a triplet of randomly chosen shapes from the same category to one of two groups – six random shapes from the same category or six random shapes from another randomly chosen category (see SI, Fig. S5B). Sample size was calculated assuming medium effect size, and was sufficient for statistical power exceeding 0.95 assuming mu0=0.5, mu1=0.7, sigma=0.05, alpha=0.05. We created for each of the largest nine categories a set of 40 such questions for a total of 360 unique questions. Groups of at least 10 participants responded to one of eight blocks of 45 questions posed in random order. See SI, Section S7 for further details.

**Interleaved meaning categories**

To test whether meaning categories are separated or interleaved, we computed the shortest path between all pairs of gallery shapes within a category, and between all pairs in different categories (total of N=653 shapes yielding ~212K pairs). See SI, Section S8 for further details. We find that the two distributions of path lengths are very similar, suggesting an interleaved geometry (between categories: Median=4 steps, 95% CI=[4,4], within categories: Median=4 steps, 95% CI=[4,4], effect= 0.06, see SI, Section S8 and Fig. S7). Second, we enumerated the number of potential gallery shapes that could be reached in K steps from a given gallery shape S. We computed the ratio between potential shapes that belong to the same category as S, and potential shapes that belong to all other categories. We find that the probability of choosing a next gallery shape that belongs to the same category, for a given number of steps, is about 1:4 (Median= 0.25, 95% CI = [0.24,0.26], see SI, Section S8 and Fig. S7). We conclude that players do not stay within the same category in a given exploitation phase only because of the proximity of these shapes.